\documentclass[aps,showpacs,superscriptaddress,preprint]{revtex4}%
\usepackage{amsfonts}
\usepackage{amsmath}
\usepackage{amssymb}
\usepackage{graphicx}%
\providecommand{\U}[1]{\protect\rule{.1in}{.1in}}

\begin{document}
\title{First-principles study of temperature-dependent diffusion coefficients:
Hydrogen, deuterium, and tritium in $\alpha$-Ti}
\author{Yong Lu}
\affiliation{LCP, Institute of Applied Physics and Computational Mathematics, Beijing
100088, People's Republic of China}
\author{Ping Zhang}
\thanks{Author to whom correspondence should be addressed. E-mail: zhang\_ping@iapcm.ac.cn}
\affiliation{LCP, Institute of Applied Physics and Computational Mathematics, Beijing
100088, People's Republic of China}
\affiliation{Beijing Computational Science Research Center, Beijing 100084, People's
Republic of China}

\begin{abstract}
We report the prediction of temperature-dependent diffusion coefficients of
interstitial hydrogen, deuterium, and tritium atoms in $\alpha$-Ti using
transition state theory. The microscopic parameters in the pre-factor and
activation energy of the impurity diffusion coefficients are obtained from
first-principles total energy and phonon calculations including the full
coupling between the vibrational modes of the diffusing atom with the host
lattice. The dual occupancy case of impurity atom in the hcp matrix are
considered, and four diffusion paths are combined to obtain the final
diffusion coefficients. The calculated diffusion parameters show good
agreement with experiments. Our numerical results indicate that the diffusions
of deuterium and tritium atoms are slower than that of the hydrogen atom at
temperatures above 425 K and 390 K, respectively.

\end{abstract}

\maketitle

Most metals and alloys are susceptible to hydrogen damage, including phase
transformations, hydrogen embrittlement, and corrosion. Diffusion of hydrogen
in metals provides an intriguing and challenging intellectual problem in
addition to the practical impact. Taking into consideration the importance of
diffusion process, many experimental measurements have been performed
\cite{Wasilewski,Papazoglou,Johnson, Volkl,Ebisuzaki, Katz}. Also, there have
been many theoretical efforts to gain diffusion coefficients for multiple
materials and the corresponding diffusion mechanisms using fundamental
electronic or atomistic approaches \cite{Wimmer, Mantina, Mantina2, Huang}. In
fact, although the diffusion experiments can determine the overall diffusion
coefficients, they cannot generally determine the microscopic physical
processes involved in the diffusion steps which are quite important for basic
understanding and practical applications. Therefore, first-principles
calculations, which have been widely used in the study of solid-state
diffusion, can help to track the microscopic diffusion processes and provide
specific quantitative values, such as formation and migration energies,
involved in the diffusion process.

Metal titanium (Ti), which combines high strength and low density, is a
suitable material for the application in aircraft construction and aerospace
engineering, as well as in the chemical industry. In most cases, titanium and
its alloys display excellent resistance to damage. However, due to the strong
affinity between titanium and hydrogen that results in a severe deterioration
in the mechanical properties, considerable restrictions exist in their
applicability under hydrogen-containing environments, where the hydrogen can
be supplied by a number of sources, including water vapor, picking acids, and
hydrocarbons. Despite the extensive work on diffusion, there is a serious lack
of reliable experimental diffusion data on titanium and many other important
systems. It is well known that metals of the IV group exhibit phase transition
from $\alpha$-phase (hcp structure) to $\beta$-phase (bcc structure) at the
reaction temperature, which is 1155 K for Ti. This relatively low transition
temperature for Ti makes diffusion experiments difficult to be solely carried
within the $\alpha$-phase. To date, the experimental diffusion coefficients of
hydrogen in $\alpha$-Ti are scattered over many orders of magnitude
\cite{Wasilewski,Papazoglou, Johnson}, while to our knowledge, systematic
\textit{ab initio} studies of hydrogen diffusion in Ti is still lacking in the literature.

The purpose of the present work is to comprehensively investigate the atomic
diffusion mechanism of hydrogen and its isotopes (deuterium and tritium) in
$\alpha$-Ti, and to obtain the specific values of the corresponding energy
barriers, vibrational free energies, and diffusion coefficients from first
principles. In the most general form, the diffusion coefficient is expressed
as
\begin{equation}
D=D_{0}e^{-Q/kT},
\end{equation}
where $Q$ is the activation energy, $D_{0}$ is a pre-factor, and $k$ is the
Boltzmann constant. According to the transition state theory (TST)
\cite{Eyring,Vineyard}, the jump rate is written as
\begin{equation}
\omega=\frac{kT}{h}\frac{Z_{TS}}{Z_{0}}e^{-\Delta H_{m}/kT},
\end{equation}
where $Z_{TS}$ and $Z_{0}$ are partition functions for the transition state
and the ground state, respectively, $h$ is the Planck's constant, and $\Delta
H_{m}$ is the enthalpy difference of migration between the transition state
and the ground state including the thermal electronic contribution. Within the
framework of the harmonic approximation, the quantum mechanical partition
functions in Eq. (2) can be expressed as
\begin{equation}
\omega=\frac{kT}{h}\frac{\prod_{i=1}^{3N-6}\left[  2\sinh\left(  \frac
{h\nu_{i}^{0}}{2k_{B}T}\right)  \right]  }{\prod_{i=1}^{3N-7}\left[
2\sinh\left(  \frac{h\nu_{i}^{TS}}{2k_{B}T}\right)  \right]  }e^{-\Delta
H_{m}/k_{B}T},
\end{equation}
where $\nu_{i}^{TS}$ and $\nu_{i}^{0}$ are the vibrational frequencies at the
transition state and the ground state, respectively. Using the expression of
the phonon free energy
\begin{equation}
F_{vib}=-kT\ln Z_{vib}=kT\int_{0}^{\infty}g(\nu)\ln\left[  2\sinh\left(
\frac{h\nu}{2kT}\right)  \right]  d\nu,
\end{equation}
the jump rate can be simply expressed as
\begin{equation}
\omega=\frac{kT}{h}e^{-\Delta F_{vib}/kT}e^{-\Delta H_{m}/kT},
\end{equation}
where the zero-point energy is included in the $F_{vib}$ term.

As depicted in Fig. 1, there are two stable interstitial positions, i.e., the
tetrahedral (T) site and octahedral (O) site, respectively. In the present
work we focus on the dual occupancy case. Thus, four jump frequencies are to
be considered. In the hcp-structured $\alpha$-Ti, there are two kinds of
impurity jumps, viz. one perpendicular to $c$ axis and the other along the $c$
axis, resulting in two diffusion coefficients which can be expressed as
\cite{Ishioka}
\begin{equation}
D_{\perp}=\frac{\omega_{\text{TO}}\omega_{\text{OT}}}{\omega_{\text{TO}%
}+2\omega_{\text{OT}}}a^{2},
\end{equation}
and
\begin{equation}
D_{\parallel}=\frac{\omega_{\text{TO}}(3\omega_{\text{OO}}\omega_{\text{TO}%
}+2\omega_{\text{OO}}\omega_{\text{TT}}+3\omega_{\text{TT}}\omega_{\text{OT}%
})}{4(2\omega_{\text{TT}}+3\omega_{\text{TO}})(\omega_{\text{TO}}%
+2\omega_{\text{OT}})}c^{2},
\end{equation}
where $\omega_{\text{TO}}$, $\omega_{\text{OT}}$, $\omega_{\text{OO}}$, and
$\omega_{\text{TT}}$ are respectively the jump rates along the four paths
shown in Fig. 1.

The density functional theory (DFT) calculations are carried out using the
Vienna \emph{ab-initio} simulation package (VASP) \cite{Kresse, Kresse2} with
the projector-augmented-wave (PAW) potential method \cite{Blochl}. The cutoff
energy for the plane-wave basis set is 450 eV. The exchange and correlation
effects are described by the generalized gradient approximation (GGA) in the
Perdew-Burke-Ernzerhof (PBE) form \cite{PBE}. We employ a $3\times3\times2$
$\alpha$-Ti supercell containing 36 host atoms to simulate hydrogen migration
in the $\alpha$-Ti matrix. To check the convergence of the formation and
migration energies, we have also considered a $4\times4\times3$ $\alpha$-Ti
supercell containing 96 host atoms and one hydrogen atom. The integration over
the Brillouin zone is carried out on $3\times3\times3$ k-point meshes
generated using the Monkhorst-Pack \cite{Monkhorst} method, which proves to be
sufficient for energy convergence of less than 1.0$\times$10$^{-4}$ eV per
atom. During the supercell calculations, the shape and size of the supercell
are fixed while all the ions are free to relax until the forces on them are
less than 0.01 eV {\AA }$^{-1}$. The phonon spectra are carried out using the
density functional perturbation theory.

\begin{figure}[ptb]
\includegraphics[width=0.8\textwidth]{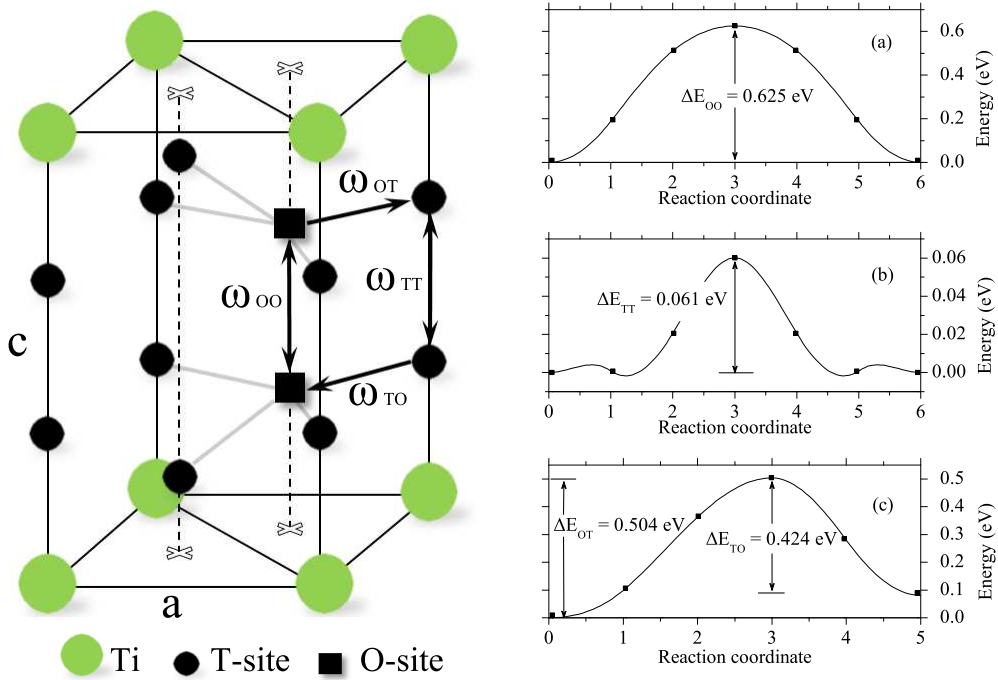}\caption{The arrangement of
interstices in the hcp lattice (left panel) and the energy profiles for a
hydrogen atom to diffuse along (a) O$\rightarrow$O, (b) T$\rightarrow$T, and
(c) O$\rightarrow$T and T$\rightarrow$O paths.}%
\label{fig1}%
\end{figure}

\begin{figure}[ptb]
\includegraphics[width=0.8\textwidth]{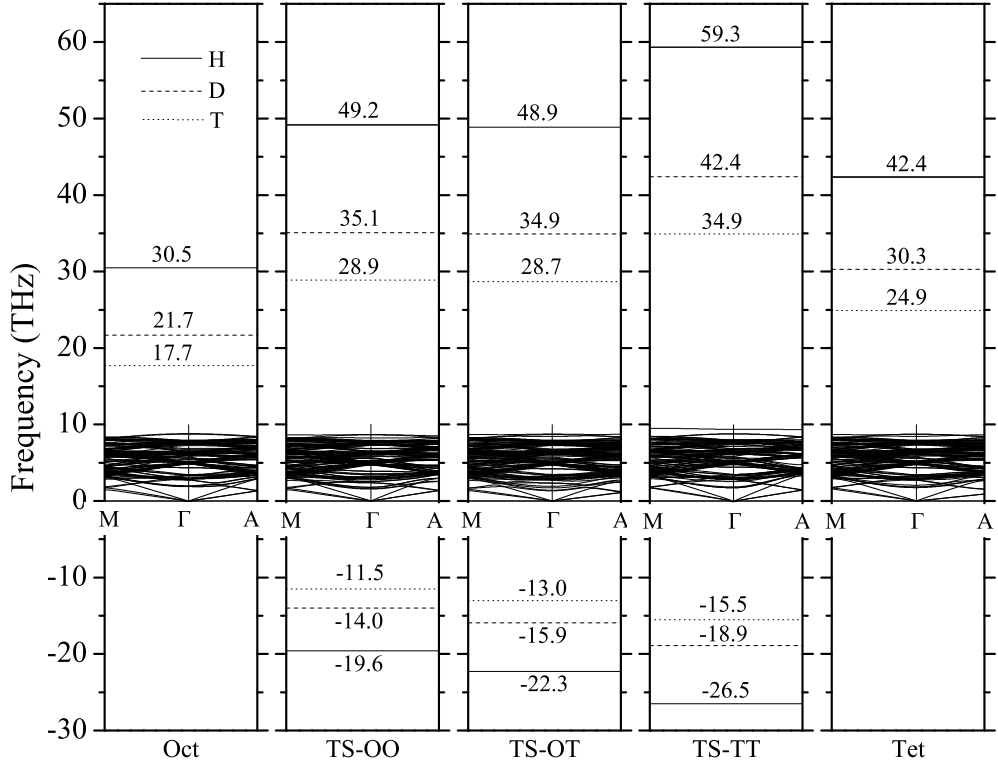}\caption{Calculated phonon
dispersions of a 36-atom $\alpha$-Ti supercell with hydrogen, deuterium, and
tritium atoms in the stable octahedral site (Oct), the transition states along
the O$\rightarrow$O (TS-OO), O$\rightarrow$T (TS-OT), and T$\rightarrow$T
(TT-TS) paths, and the metastable tetrahedral site (Tet), respectively. The
modes with imaginary frequencies at the transition states correspond to the
motion of the hydrogen, deuterim, and tritium atoms across the barriers.}%
\label{fig2}%
\end{figure}

The theoretical equilibrium lattice parameters $a$ and $c/a$ for $\alpha$-Ti
are 2.939 \r{A}$~$ and 1.583 \r{A}, respectively, in accordance with the
experimental data of 2.950-2.957 \r{A}$~$and 1.585-1.587
\cite{Vohra,Ostanin,Zhang}. By comparing the formation energies of a hydrogen
defect in O-site and T-site, we find that the O-site is more favorable in
energy, lower than the T-site with a 36-atom supercell by 0.080 eV. The
96-atom supercell gives the essentially same result with an energy difference
of 0.082 eV. For the calculation of migration energies and phonon frequencies,
an important issue concerns the location of the saddle-point position for
hydrogen atom. Here, each saddle-point structure and the associated
minimum-energy pathway (MEP) were calculated by employing the climbing image
nudged elastic band (CINEB) method \cite{CINEB}. As displayed in Fig.
1(a)-(c), when hydrogen atom diffuses from the stable site to the transition
site (TS), the energy profile can be well described by a sinusoidal curve, as
originally suggested by Wert and Zener \cite{Wert}. In the case of
O$\rightarrow$O path and T$\rightarrow$T path, the energy is found to display
a single maximum, corresponding to a saddle point at the high-symmetry
position located half way between neighboring sites. When hydrogen atom
diffuses along the O$\rightarrow$T path or T$\rightarrow$O path, the
transition state locates close to the T site, and the distance between the
saddle point and T-site (O-site) is 0.852 \r{A}$~$(1.001 \r{A}). In Table I,
we listed the migration energies of the four diffusion paths using 36- and
96-atom models, respectively. By comparison of the two models, one can see
that a 36-atom supercell is already convergent. Actually, the difference in
migration energy of these two models are 0.04 eV at most for O$\rightarrow$O
path, indicating that a 36-atom supercell can be reliably used in the
simulations of the single hydrogen difusion energetics. Thus, in the follows
we adopt a 36-atom supercell to perform the phonon dispersion calculations.

\begin{table}[ptb]
\caption{Calculated migration enthalpy difference $\Delta H_{m}$ of atomic
hydrogen diffusion via O$\rightarrow$O path, T$\rightarrow$T path,
O$\rightarrow$T path, and T$\rightarrow$O path for $3\times3\times2$ and
$4\times4\times3$ hcp Ti supercell models, respectively.}%
\begin{ruledtabular}
\begin{tabular}{cccccccccccccccc}
&$3\times3\times2$&&&&$4\times4\times3$\\
\cline{2-9}
&O-O&T-T&O-T&T-O&O-O&T-T&O-T&T-O\\
\hline
$\Delta H_m$&0.662&0.042&0.525&0.439&0.625&0.061&0.504&0.424\\
\end{tabular} \label{a}
\end{ruledtabular}
\end{table}

The phonon dispersions of the 36-atom supercell with a hydrogen atom at the
O-site, T-site, and three TS sites are shown in Fig. 2. When a hydrogen atom
occupies an octahedral site, its phonons show a threefold-degenerate
dispersionless branch at 30.5 THz. At the T-site, the frequencies of the
hydrogen atom show a similar character with respect to those in the O-site,
and the threefold-degenerate branch moves to 42.4 THz. In both cases, the
frequencies are positive, indicating a true minimum in the energy. When the
hydrogen atom moves to a transition state, its phonon branches, as shown in
Fig. 2, are split into a degenerate doublet, which shift upwards to some
extent, and a non-degenerate singlet, which shifts downwards so as to be
imaginary, implying the spontaneous diffusion instability for the defective
hydrogen in the transition state. By definition, the transition state is
characterized by the occurrence of one negative eigenvalue in the dynamical
matrix. Here the doublet modes at positive frequency correspond to the
vibrations of the hydrogen atom perpendicular to the diffusion path, while the
singlet mode at imaginary frequency corresponds to the motion of the hydrogen
atom along its diffusion path.

When the hydrogen defect locates at the transition state along the
T$\rightarrow$T path, we note that the hydrogen-related branch has the highest
frequency of 59.3 THz compared to 49.2 THz in O$\rightarrow$O path and 48.9
THz in O$\rightarrow$T path. In fact, the transition site along the
T$\rightarrow$T path locates at the geometric center of triangle formed by its
three nearest Ti atoms, thus the interstitial space is much more confined with
respect to the other two transition states. Besides, there is a phonon branch
between 9.0 and 10.0 THz in the transition state along the T$\rightarrow$T
path, which is obviously off from the bulk of Ti-related phonon branches. This
branch is related to motions of Ti atoms, which couple with motions of the H
impurity. As we expect, the impurity atom related frequencies decrease with
increasing mass from hydrogen to its isotopes (cf. Fig. 2).

\begin{figure}[ptb]
\includegraphics[width=0.8\textwidth]{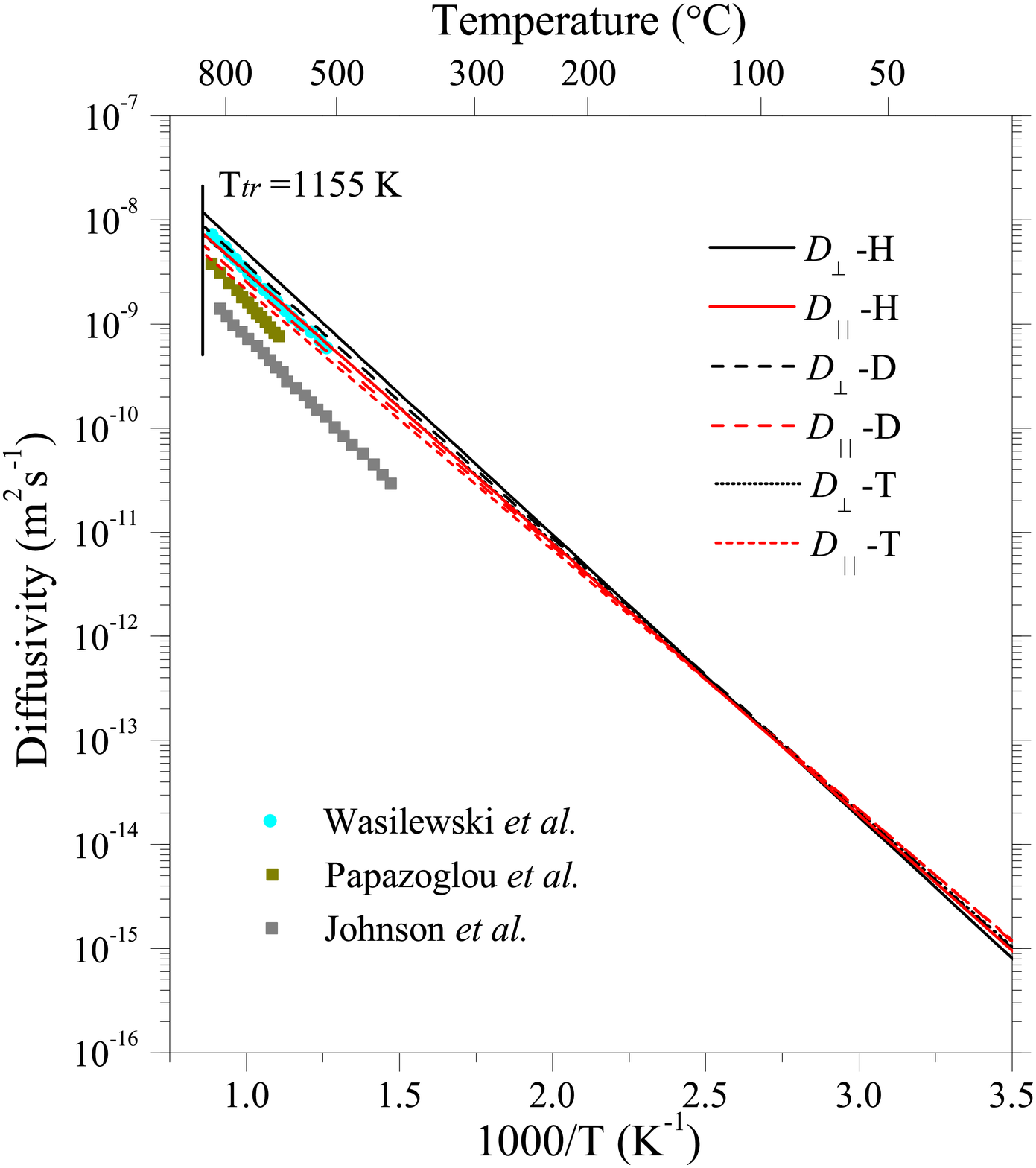}\caption{(Color online)
Calculated and experimental diffusion coefficients of atomic hydrogen,
deuterium, and tritium in $\alpha$-Ti. The T$_{tr}$=1155 K stands for the
phase transition temperature of metal Ti from $\alpha$-phase to $\beta
$-phase.}%
\label{fig3}%
\end{figure}

\begin{figure}[ptb]
\includegraphics[width=0.8\textwidth]{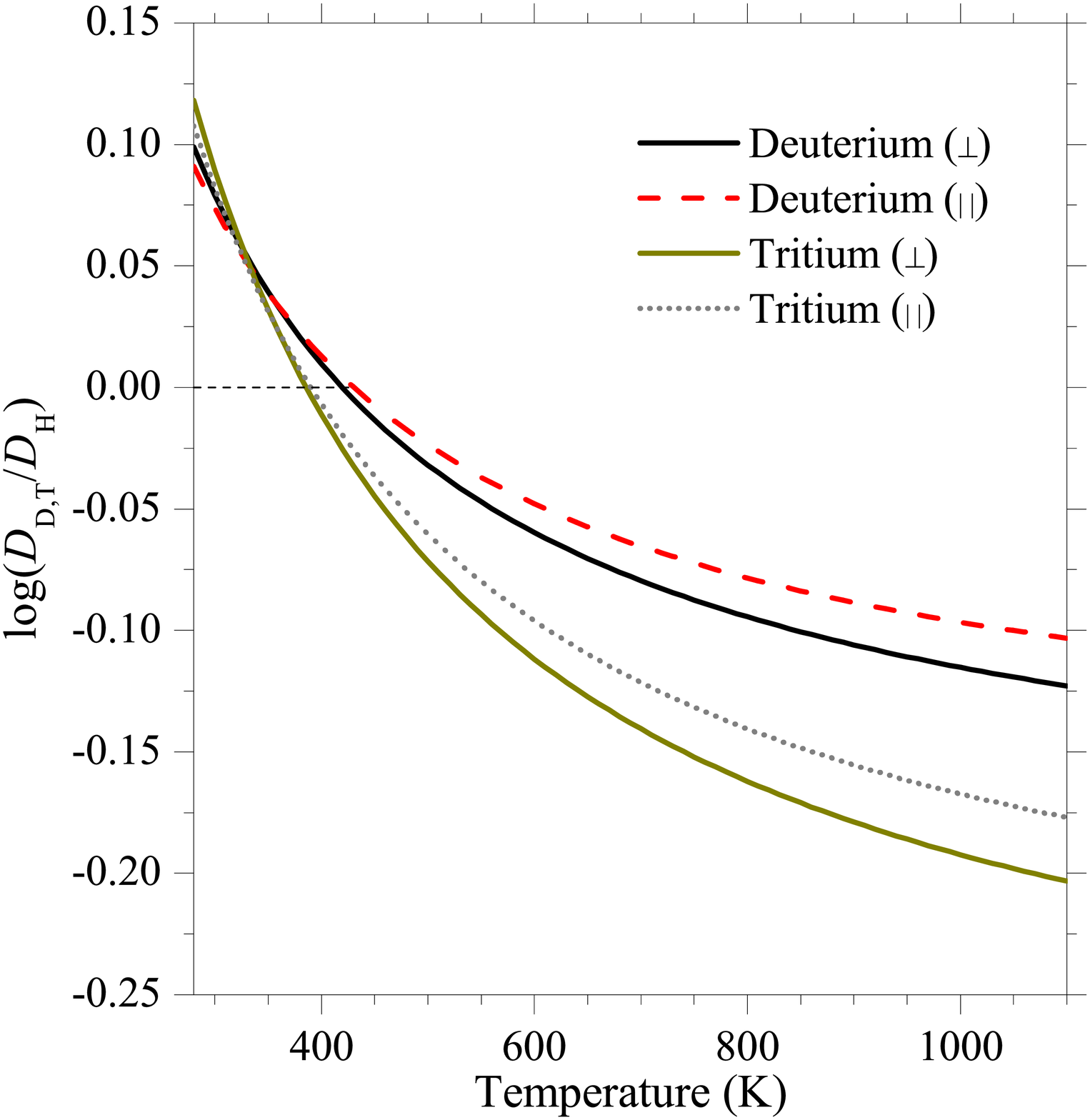}\caption{(Color online) The
ratios of the diffusivity of deuterium and tritium to hydrogen atoms. ($\bot$)
and ($\parallel$) represent the impurity diffusions perpendicular to and
parallel to $c$ axis, respectively.}%
\label{fig4}%
\end{figure}\begin{table}[ptbptb]
\caption{Calculated diffusion pre-factors ($D_{0\perp}$ and $D_{0\parallel}$)
and activation energy ($Q_{\bot}$ and $Q_{\parallel}$) for atomic hydrogen,
deuterium, and tritium in $\alpha$-Ti. The temperatures represent the ranges
over which diffusion coefficients were fit to extract corresponding
pre-factors and activation energies. For comparison, experimental results are
also listed.}%
\begin{ruledtabular}
\begin{tabular}{cccccccccccccccc}
Method&$D_{0\perp}$ (m$^{2}$/s)&$D_{0\parallel}$ (m$^{2}$/s)&$Q_{\bot}$ (kJ/mol)&$Q_{\parallel}$ (kJ/mol)&T (K)\\
\hline
Hydrogen&2.506$\times$10$^{-6}$&1.275$\times$10$^{-6}$&51.90&49.89&280-1000\\
Deuterium&1.587$\times$10$^{-6}$&8.630$\times$10$^{-7}$&50.30&48.50&280-1000\\
Tritium&1.219$\times$10$^{-6}$&6.781$\times$10$^{-7}$&49.59&47.85&280-1000\\
Expt.$^{a}$&1.672$\times$10$^{-6}$&&51.80&&773-1097\\
Expt.$^{b}$&3.057$\times$10$^{-6}$&&62.46&&884-1102\\
\end{tabular} \label{a}
$^{a}$ Reference \cite{Wasilewski}, $^{b}$ Reference
\cite{Papazoglou}
\end{ruledtabular}
\end{table}

Integration of the phonon dispersions over the entire Brillouin zone allows to
compute the temperature-dependent enthalpy, entropy, and free energy, which
can be used to evaluate the diffusion coefficient according to Eq. (1). In
Fig. 3 we showed the Arrhenius plot of the computed diffusion coefficients
together with experimental values in Refs.
\cite{Wasilewski,Papazoglou,Johnson} for comparison. The diffusion coefficient
curves for $D_{\perp}$ and $D_{\parallel}$ are almost indistinguishable below
500 K. With increasing temperature, the value of $D_{\perp}$ becomes somewhat
higher than $D_{\parallel}$. Both of these two diffusion coefficients are in
agreement with the experimental data by Wasilewski \emph{et al.}
\cite{Wasilewski} and Papazoglou \emph{et al.} \cite{Papazoglou} (cf. Fig. 3).
By linear fitting of the diffusion coefficient curves, we obtained the
diffusion pre-factors ($D_{0\perp}$ and $D_{0\parallel}$) and activation
energies ($Q_{\perp}$ and $Q_{\parallel}$) according to the Arrhenius plot of
Eq. (1), i.e., $\ln(D)=\ln(D_{0})-Q/kT$. In Table 2 we listed our fitting
results of pre-factors and activation energies. A linear fit between 280 K and
1000 K of the calculated data for hydrogen gives $Q_{\perp}$=51.9 kJ/mol and
$Q_{\parallel}$=49.89 kJ/mol, in agreement with experimental data ranging from
51.80 kJ/mol to 62.46 kJ/mol \cite{Wasilewski, Papazoglou}. The pre-factors
are $D_{0\perp}$=2.506 $\times$10$^{-6}$ m$^{2}$/s and $D_{0\parallel}%
$=1.275$\times$10$^{-6}$ m$^{2}$/s, respectively. One can see that
$D_{0\parallel}$ is nearly the half of $D_{0\perp}$. The value of $D_{0\perp}$
is well located in the experimental range from 1.672$\times$10$^{-6}$ m$^{2}%
$/s to 3.057$\times$10$^{-6}$ m$^{2}$/s.

The temperature-dependent diffusion coefficients for the isotopes deuterium
and tritium in $\alpha$-titanium can be calculated based on the electronic
structure results of hydrogen in titanium. One would expect a lower
diffusivity for the heavier isotopes, since the impurity-related frequencies
decrease with increasing mass (Fig. 2). In Fig. 4 we depicted the ratio of the
diffusivity of deuterium and tritium to hydrogen. One can see clearly that
this is only the case at temperatures above 425 K for deuterium and 390 K for
tritium. At lower temperatures, the ordering is reversed. The reason is that
at low temperatures, the zero-point energy effect is more pronounced \cite{Wimmer}, which
increases the jump rate through an exponential dependence, since the energy
difference between the transition and ground states shows a reduction trend
from hydrogen to its isotopes. As the temperature increases, the zero-point
energy effect is offset due to the lower frequencies of isotopes. At 425 K,
these two effects achieve a balance for deuterium. Also, a crossover
temperature of 390 K is predicted for tritium, as depicted in Fig. 4. 

In summary, we have systematically studied the temperature-dependent diffusion
coefficients of atomic hydrogen and its isotopes in $\alpha$-Ti using
transition state theory with accurate first-principle total energy and phonon
calculations. The quantitative agreement between the calculated and
experimental diffusion coefficients has been achieved, which demonstrates the
validity and practicability of the present theoretical method. Our calculated
ratios of the diffusivity of isotopes and hydrogen provide a good reference
for future experimental measurements on $\alpha$-Ti.

This work was supported by NSFC under Grant No. 51071032 and by CAEP
Foundations for Development of Science and Technology under Grant No. 2011A0301016.

\end{document}